\documentclass[aps,prd,twocolumn,preprintnumbers,superscriptaddress,floatfix]{revtex4}

\usepackage{multirow, graphicx,amssymb,url,mathrsfs,amsmath}
\usepackage{eucal,wrapfig,boxedminipage,setspace,subfigure}
\usepackage{amsxtra,amstext,latexsym,dsfont}

\def\IR{{\hbox{{\rm I}\kern-.2em\hbox{\rm R}}}}
\def\IB{{\hbox{{\rm I}\kern-.2em\hbox{\rm B}}}}
\def\IN{{\hbox{{\rm I}\kern-.2em\hbox{\rm N}}}}
\def\IC{\,\,{\hbox{{\rm I}\kern-.59em\hbox{\bf C}}}}
\def\IZ{{\hbox{{\rm Z}\kern-.4em\hbox{\rm Z}}}}
\def\IP{{\hbox{{\rm I}\kern-.2em\hbox{\rm P}}}}
\def\IH{{\hbox{{\rm I}\kern-.4em\hbox{\rm H}}}}
\def\ID{{\hbox{{\rm I}\kern-.2em\hbox{\rm D}}}}

\newcommand{\beq}{\begin{equation}}
\newcommand{\eeq}{\end{equation}}
\newcommand{\bea}{\begin{eqnarray}}
\newcommand{\eea}{\end{eqnarray}}

\begin{document}

\voffset 1cm

\newcommand\sect[1]{\emph{#1}---}

\title{Mass hierarchies in gauge theory with two index symmetric representation matter}

\author{Anja Alfano}
\affiliation{ STAG Research Centre \&  Physics and Astronomy, University of
Southampton, Southampton, SO17 1BJ, UK}

\author{Nick Evans}
\affiliation{ STAG Research Centre \&  Physics and Astronomy, University of
Southampton, Southampton, SO17 1BJ, UK}

\begin{abstract}
In gauge theories, the running of the anomalous dimension of a fermion bilinear operator is believed to lead to chiral symmetry breaking when $\gamma=1$. Naïvely using perturbative results to judge when $\gamma=1$ leads to the possibility of large dynamically generated mass hierarchies in models with both two index symmetric representation and fundamental representation fermions.  In this paper we study a holographic model of this physics to predict the separation in scales in the meson spectrum. We study $SU(N_c)$ theories at different $N_c$ with one flavour of two index symmetric representation as a function of the number of fundamental fermion flavours $N_f^F$. The largest hierarchy we find is for $N_c=7$ and $N_f^F=22$ where the rho mesons made of the two different representations are separated in scale by a factor of 13. For general $N_c$ the hierarchy can be made greater than 7 by tuning $N_f^F$.  We display the hierarchy as a function of $N_f^F$ and investigate the quark mass dependence. These predictions do depend on the extrapolation of $\gamma$ from the perturbative regime - even in a pessimistic scenario a distinct gap  can be achieved.
\end{abstract}%

\maketitle

\newpage

\section{Introduction}\vspace{-0.5cm}

In QCD the logarithmic running of the asymptotically free gauge coupling leads to an infra-red strong coupling regime in which chiral symmetry is dynamically broken, generating the scale of the baryon masses. QCD also exhibits confinement at the same or very similar scale (although there is no formal order parameter for confinement in the light quark theory). These same behaviours are expected in many asymptotically free non-supersymmetric gauge theories with fermionic matter and, for example, lattice studies have confirmed this for theories: with $N_f^F$ fundamentals other than $N_f^F=3$ \cite{LSD:2014nmn,LatticeStrongDynamics:2018hun}; fermions in the adjoint \cite{Bergner:2022snd}; two index symmetric representation \cite{Fodor:2015zna}; and two index anti-symmetric representation (with fundamentals) \cite{Ayyar:2018zuk}. 

It is an interesting question as to whether such theories can generate multiple scales \cite{Marciano:1980zf,Evans:2020ztq}. A natural way to do this is  to include fermions in two (or more) representations which might interact with the strongly coupled glue differently. Were higher dimension representations to condense at a lower coupling value a gap to the confinement scale might also be possible.

A criteria for the formation of a chiral symmetry breaking bi-fermionic operator (with perturbative dimension $\Delta=3$) has been deduced from both gap equation analysis \cite{Cohen:1988sq} and holographic models \cite{Jarvinen:2011qe,Kutasov:2011fr,Alvares:2012kr}. It is:  when the anomalous dimension of the bi-fermion operator, $\gamma$, rises to one, so that the mass and operator are of equal dimension $\Delta=2$, there is an instability to condensation. This is most simply seen in holography where a bulk scalar in AdS$_5$ (with unit radius), dual to an operator has mass squared $M^2=\Delta(\Delta-4)$ \cite{Witten:1998qj}
and the Breitenlohner-Freedman stability bound in AdS$_5$ is $M^2=-4$ which is precisely saturated at $\Delta=2$  \cite{Breitenlohner:1982jf}. 

In \cite{Evans:2020ztq} we performed a simple analysis of using perturbative results, extrapolated into the non-perturbative regime. We estimated the scales where the $\gamma=1$ criteria occurred for different multi-representation theories. Of course the results depend on the extrapolation - there, and here, we follow the ansatz of \cite{Appelquist:1996dq} which for example places the edge of the conformal window with fundamentals at approximately $4N_c$. Lattice results suggest it may be as low as $N_f^F=10$ in the SU(3) gauge theory \cite{Hasenfratz:2023wbr} but the ansatz we take provides some sensible estimate for the running with the associated physical flavour numbers adjustable by the readers expectations for where the conformal windows begin. We will also use a different ansatz where the anomalous mass of the AdS scalar,$\Delta m^2$, is double the perturbative estimate to provide a most pessimistic scenario.  

In a theory with a single Dirac fermion in the two-index symmetric representation and $N_f^F$ Dirac fermions in the fundamental representation it was possible to find theories   where the separation of scales was possibly of order ten. These theories are therefore of potential interest and also look open to lattice analysis. The most extreme cases are quite walking theories and this might present problems on the lattice.

In this paper, therefore, we perform a further analysis where we study the meson spectrum of these theories ($SU(N_c)$ gauge theory with one two index symmetric representation fermion and $N_f^F$ fundamentals) as a function of $N_c$ and $N_f^F$. Our goal is to go beyond the naïve perturbative analysis by including the dynamics of chiral symmetry breaking, not just estimating the scale where the instability sets in. We are also interested in how finely tuned $N_f^F$ needs to be for the gap to be explorable. For example, the lattice can study fermions in multiples of four more easily than arbitrary choices. Reaching the massless limit can also be challenging on the lattice as we will discuss.  

Our tool is a holographic model of the dynamics \cite{Alho:2013dka}. We have previously used this model to study walking dynamics \cite{Erdmenger:2014fxa,Belyaev:2018jse} and composite higgs models \cite{Erdmenger:2020lvq,Erdmenger:2020flu,Erdmenger:2024dxf}. The model is based on the D3/probe D7 system \cite{Karch:2002sh,Kruczenski:2003be,Erdmenger:2007cm}
in holography. It is a simple system in which one uses the weakly coupled Dirac-Born-Infeld (DBI) action to study the position and fluctuations of the D7 branes in AdS$_5$. Remarkably this is believed to be an exact study of quarks in the ${\cal N}=2$ supersymmetric gauge theory with ${\cal N}=4$ glue. Further one can perturb this system, breaking supersymmetry and seeing chiral symmetry breaking (for example by adding a magnetic field \cite{Albash:2007bk} or by deforming the background geometry \cite{Babington:2003vm}). From the perspective of the DBI action these changes manifest in the  mass squared for the scalar dual to the quark bilinear (the position of the D7 brane). That mass squared becomes dependent on the radial coordinate in AdS, dual to the renormalization group  scale \cite{Alvares:2012kr}. The duality translates this to a running dimension of the operator and condensation is triggered when $\gamma=1$. Our AdS/Yang-Mills model \cite{Alho:2013dka} simply takes this precise system and imposes the running of a gauge theory to determine the dynamics.  Here we use the perturbative two loop results for the running of $\alpha$ extended into the non-perturbative regime as our ansatz for this running. Whilst this may be insecure we can at least see the response of the theory to different runnings and critical couplings. 

The main conclusions, as we will see, are that the holographic model does support the presence of the gaps seen in the perturbative analysis (in part of course because it incorporates those runnings). These gaps can be larger than an order of magnitude: for example in  SU(4) theory with $N_f^F=11$, SU(5) with $N_f^F=15$,  and SU(7) with $N_f^F=22$. We caveat this with a more pessimistic translation of the running coupling to the AdS scalar mass correction, $\Delta m^2$, but even there the gaps can be a factor of 2 which is still a clear scale separation. We show the theories with large gaps are quite slowly running and live close to the fixed points which only just trigger chiral symmetry breaking. We also identify theories with smaller gaps and more QCD-like running. We hope to inspire lattice simulations of these theories which can show from first principles a separation between chiral symmetry breaking scales and even the confinement scale.   

\section{The Holographic Model}

The holographic AdS/Yang-Mills model was first developed in \cite{Alho:2013dka}. Here we briefly present the action and equations needed to compute the meson masses and decay constants. 

The action for the fields for the two representations in the model is
\begin{multline} \label{eq: general_action}
S_{boson} = \sum_R \int d^5 x ~ \rho^3 \left( \frac{1}{r^2} (D^M X_R)^{\dagger} (D_M X_R)\right. \\
    \left. + \frac{\Delta m_R^2}{\rho^2} |X_R|^2 + \frac{1}{2 g_{R5}^2} \left(\vphantom{\frac{1}{2}} F_{R,MN}F_{R}^{MN} + (V \leftrightarrow A) \vphantom{\frac{1}{2}} \right) \right) \, .
\end{multline}
Note the only interplay between the two representations is through their contributions to $\Delta m^2_R$.
The five-dimensional coupling may be obtained by matching to the UV vector-vector correlator \cite{Erlich:2005qh,DaRold:2005mxj}, and  is given by 
\begin{equation}
g_{R5}^2 = \frac{12 \pi^2}{d(R)~N_{f}(R)} \, ,
\end{equation}
where $d(R)$ is the dimension of the fermion's representation and $N_f(R)$ is the number of Dirac flavours in that representation.

The metric for each representation is a five-dimensional asymptotically AdS  spacetime
\begin{align}  \label{thing2}
ds_R^2 = r^2 dx^2_{(1,3)} + \frac{d \rho^2}{r^2} \, ,
\end{align} 
with $r^2 = \rho^2 +|X_R|^2$ the holographic radial direction corresponding to the energy scale, and with the AdS radius set to one.

The dynamics of a particular gauge theory, including quark contributions to any running coupling, are included through the choices of $\Delta m_R^2$ in \eqref{eq: general_action}. Our starting point is the perturbative result for the running of $\gamma$. Expanding $M^2=\Delta(\Delta-4)$ at small $\gamma$ gives
\begin{align} \label{dm}
\Delta m^2 = - 2 \gamma.
\end{align}
Since  the true running of $\gamma$  is not known non-perturbatively,  we extend the perturbative results as a function of renormalization group (RG) scale $\mu$ to the non-perturbative regime. We will directly set the field theory RG scale $\mu$ equal to the holographic RG scale $r= \sqrt{\rho^2+|X_R|^2}$. 

The two-loop result for the running coupling, $\alpha(\mu)$, in a gauge theory with multi-representational matter is 
\begin{align}
\mu \frac{d \alpha}{d \mu} = - b_0  \alpha^2 - b_1  \alpha^3 \, ,
\end{align} 
with 
\begin{equation}   \label{running}
\begin{array}{ccl}
b_0 &=& \frac{1}{6 \pi} \left(11 C_{2}(G) - 4 \sum_{R} T(R)N_f(R) \right) \, ,\\ &&\\
b_1 &=& \frac{1}{24 \pi^2} \left(34 C^2_{2}(G) \right. \\ &&\\ && \left.-\sum_{R} \left(20 C_{2}(G) + 12 C_{2}(R) \right) T(R) N_f(R) \right)  \, .
\end{array}
\end{equation}
where $T(R)$ is half the Dynkin index and $C_2(R)$ the quadratic Casimir of the representation $R$.
We use the one-loop anomalous dimension result (going beyond one loop does not introduce any new structures into what is already a guess as to the form of the running)
\begin{align} \label{grun}
\gamma = \frac{3~C_2(R)}{2 \pi}~\alpha.
\end{align}
Note these gauge theory results are all gauge invariant and scheme independent as discussed in \cite{Appelquist:1996dq,Ryttov:2010iz} in the context of the conformal window. 

A more pessimistic ansatz which we will also use below is to remove the factor of 2 in (\ref{dm}) which then requires $\alpha$ to be larger in the IR to trigger chiral symmetry breaking for the symmetric representation. This reduces the number of flavours of fundamentals one can include in the theory before one enters the conformal window. Further the speed of running in the fundamental theory below the symmetric representation dynamical mass will be faster since it itself is controlled by the value of $\alpha$ and this will tend to close the gap between the scales where the two representations condense. We will use this to show how big the errors in the analysis could potentially be.

To find the vacuum of the theory, with chiral condensates,  we set all fields to zero except for $|X_R| = L_R(\rho)$. For $\Delta m_R^2$ a constant, the equation of motion  from \eqref{eq: general_action} is
\begin{align}
\partial_{\rho} (\rho^3 \partial_{\rho} L_R) - \rho ~ \Delta m_R^2 L_R = 0 \label{eq: vacuum qcd} \, .
\end{align}
At large $\rho$, the UV,  the solution  is given  by  $L_R(\rho) = m_R + c_R/\rho^2$, with $c_R$ the fermion condensates of dimension three and $m_R$, the mass, of dimension one. We numerically solve \eqref{eq: vacuum qcd} with our input $\Delta m_R^2$ for the function $L_R(\rho)$. We use IR boundary conditions at the point where the fermions move on mass shell 
\begin{align} \label{vacIR}
L_R(\rho)|_{\rho=\rho_R^{IR}} = \rho_R^{IR} \, ,&& \partial_{\rho} L_R(\rho)|_{\rho=\rho_R^{IR}} = 0 \, .
\end{align}
We numerically vary $\rho_{IR}$ until the value of $L_R$ at  the boundary is the desired fermion mass.  We call the vacuum solutions $L_{R0}(\rho)$ with IR value $L_R^{IR}$. 

In the models we study the two index symmetric representation always condenses at a higher $\rho_R^{IR}$ than the fundamental representation. At that scale we integrate out the symmetric representation fermions and remove their contribution to the  beta function at lower scales. We show an example running of $\Delta m_R^2$ for the two representations in Figure 1. 

\noindent{\bf The meson sectors}

The mesons are fluctuations in the various fields of the model in \eqref{eq: general_action}. A fluctuation is written as $F(\rho) e^{-ik\cdot x}, ~ M^2 = -k^2$ and IR boundary conditions $F(L_R^{IR})=1, F'(L_R^{IR})=0$ are used. One seeks the values of $M^2$ where the UV solution falls to zero, so there is only a  fluctuation in the operator and not the source. 

The fluctuations of $L_R(\rho)$ give rise to scalar mesons. They are obtained by writing  $L_R= L_{R0}+S_R$  and where $r^2=\rho^2 + L_{R0}^2$.
 The equation of motion for the fluctuation reads
\begin{equation} \label{eq: eqn of motion_scalar} \begin{array}{lc}
\partial_{\rho} (\rho^3 \partial_{\rho} S_R(\rho)) - \rho (\Delta m_R^2) S_R(\rho) \\ & \\  - \rho L_{R0}(\rho) S_R(\rho) \frac{\partial \Delta m_R^2}{\partial L} |_{L_{R0}}+ M^2 \frac{\rho^3}{r^{4}} S_R(\rho) & = 0.  \end{array}
\end{equation}

The vector-mesons are obtained from fluctuations of the vector gauge field and satisfy the equation of motion
\begin{align} \label{eq: eqn of motion_vector}
\partial_{\rho} (\rho^3 \partial_{\rho}V_R(\rho)) + M^2_{V} \frac{\rho^3}{r^{4}} V_R(\rho) = 0.
\end{align}
A canonically normalized kinetic term for the vector meson requires
\begin{equation}  
\int d\rho~ \frac{\rho^3}{g_5^2 r^4} V_R^2 = 1. \label{Vnorm}
\end{equation}
\vspace{-0.4cm}

The dynamics of the axial-mesons is described by 
\begin{equation}  \label{eq: eqn of motion_axial}
\partial_{\rho} (\rho^3 \partial_{\rho} A_R(\rho)) - g_{R5}^2 \frac{\rho^3 L^2_{R0}}{r^2} A_R(\rho) + \frac{\rho^3 M^2_{A} }{r^{4}} A_R(\rho) = 0\, . 
\end{equation}

To compute decay constants, we couple the meson to an external source. Those sources are described as fluctuations with a non-normalizable UV asymptotic form. We fix the coefficient of these solutions by matching to the gauge theory in the UV.  In the UV we expect $L_{R0}(\rho) \sim 0$ and we can solve the equations of motion for the scalar, $L= K_S(\rho) e^{-i q\cdot x}$, vector $V^\mu= \epsilon^\mu K_V(\rho) e^{-i q\cdot x}$, and axial $A^\mu= \epsilon^\mu K_A(\rho) e^{-i q\cdot x}$ fields. Each satisfies the UV asymptotic equation
\begin{equation}  \label{thing}
\partial_\rho [ \rho^3 \partial_\rho K] - \frac{q^2}{\rho} K= 0\,. 
\end{equation}
with solution 
\begin{equation} \label{Ks}
K_i = N_i \left( 1 + \frac{q^2}{4 \rho^2} \ln \left(\frac{q^2}{\rho^2}\right) \right),\quad (i=S,V,A),
\end{equation}
where $N_i$ are normalization constants that are not fixed by the linearized equation of motion.
Substituting these solutions back into the action gives the scalar, vector, and axial vector correlators. Performing the usual AdS/QCD matching to the UV gauge theory  requires us to set \cite{Erlich:2005qh,Alho:2013dka} 
\begin{equation}  \label{eq: match}
N_S^2 = \frac{d(R) ~ N_f(R)}{24 \pi^2 }, \hspace{0.5cm} N_V^2 = N_A^2 = \frac{g_{R5}^2 ~ d(R) ~ N_f(R)}{24 \pi^2 }. 
\end{equation}

The vector meson decay constant is then given by the

\begin{center}
\includegraphics[width=6.7cm,height=4.8cm]{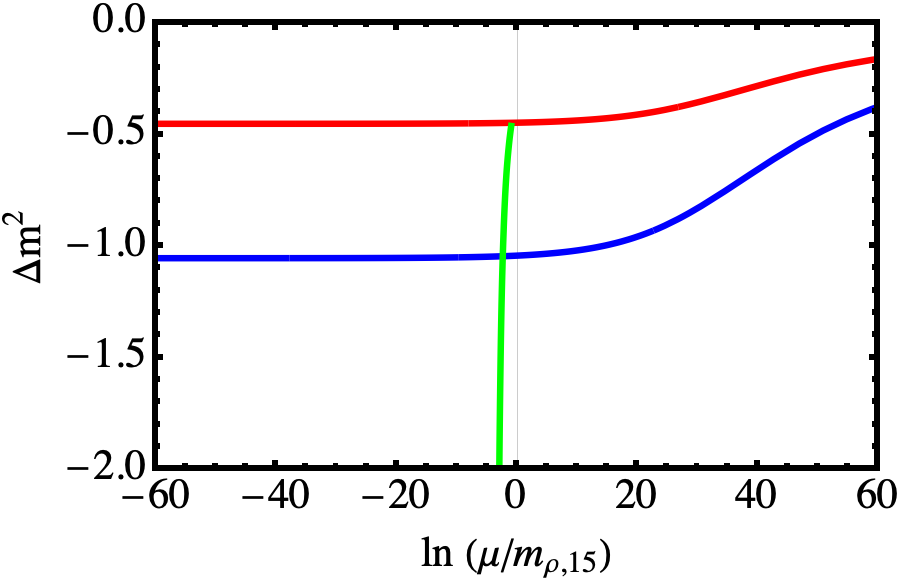}

\textit{Figure 1: SU(5) gauge theory with $N^F_f=15$: The running of $\Delta m^2$ for the two index symmeric 15 dimensional rep.  (blue), fundamental (red) and for the fundamentals with the higher dim rep. decoupled below the scale where it is  on mass shell (green). The energy scales are given in units of the $\rho$-meson mass in the $15$ sector. BF bound violation occurs at $14.2$ for the 15 and $-2.3$ for the fundamental with the 15 decoupled.}
\end{center} \vspace{-2cm}

overlap term between the meson and the external source
\begin{equation} F_{RV}^2 = \int d \rho \frac{1}{g_{R5}^2} \partial_\rho \left[- \rho^3 \partial_\rho V_R\right] K_{RV}(q^2=0)\,.
\label{rhodecay}
\end{equation}  

The axial meson normalization and decay constant are  given by  \eqref{Vnorm} and \eqref{rhodecay} with replacement $V\rightarrow A$.

The pion decay constant can be extracted since $\Pi_{AA} = f_\pi^2$, with 
\begin{equation} f_{R\pi}^2 = \int d \rho \frac{1}{g_5^2}  \partial_\rho \left[  \rho^3 \partial_\rho K_{RA}(0)\right] K_{RA}(0)\,.
\end{equation}

To compute the pion mass we work in the $A_\rho =0$ gauge and write $A_\mu = A_{\mu\perp} + \partial_\mu \phi$. The $\phi$ and $\pi$ fields (the phase of $X$)  mix to describe the pion 
\begin{equation} \label{eq: pion_full}
\begin{split}
\partial_{\rho} (\rho^3 \partial_{\rho} \phi_R (\rho))  - g_{R5}^2 \frac{\rho^3 L_{R0}^2}{r^4} (\pi_R(\rho) - \phi_R(\rho)) &=0\, , \\
q^2 \partial_{\rho} \phi_R(\rho) - g_5^2 L_{R0}^2 \partial_{\rho} \pi_R(\rho) &= 0\, .
\end{split}
\end{equation}
Substituting the lower equation of \eqref{eq: pion_full} into the upper gives
\begin{equation}
\partial_{\rho} \left( \rho^3 ~ L_{R0}^2 ~ \partial_{\rho} \pi_R \right) + M^2_{\pi} \frac{\rho^3~L_{R0}^2}{r^4} \left( \pi_R - \phi_R \right) = 0  \, .\label{eq: 4}
\end{equation}
The numerical work though is simplified if one neglects the mixing between the heavy axial sector and the light $\pi_R$ field, setting $\phi_R=0$. The results continue to be consistent with expectations such as the Gell-Mann-Oakes-Renner relation.

\begin{center}
\includegraphics[width=6.7cm,height=4.8cm]{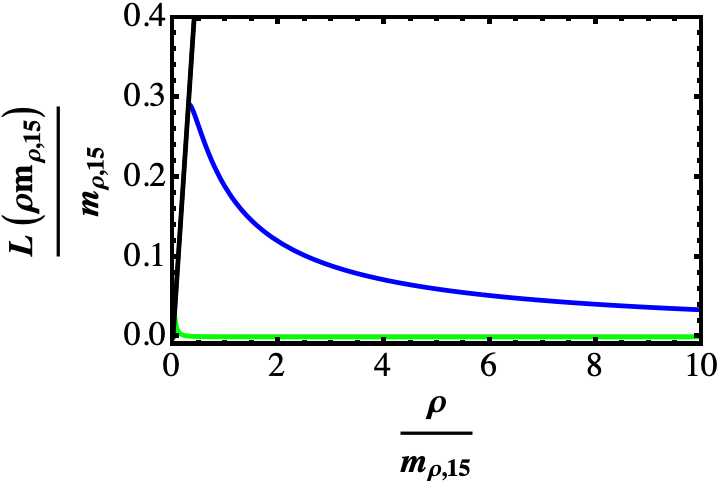}

\textit{Figure 2: SU(5) gauge theory with $N^F_f=15$: The $L_R(\rho)$ functions for the 15 (blue) and 5 (green) representations with $m_{IR,S}\approx0.29$ and $m_{IR,F}\approx0.027$ respectively in units of the $\rho$ meson mass in the 15 sector.}

\end{center} \vspace{-1cm}

\section{$N_c$ and $N_f^F$ Dependence In The Massless Theory}

We can now move to discussion of the predictions of the holographic model.  
\vspace{-0.5cm}

\subsection{$N_c=5$ Theory} \vspace{-0.5cm}

We will begin by looking at an extreme theory that encapsulates the gains and losses in these models. So first we pick the SU(5) gauge theory with one two index symmetric, dimension 15 representation and $N_f^F$ fundamental flavours.
 This model was identified as having the largest gap between the condensation scales of the two representations in \cite{Evans:2020ztq} for $N_f^F=15$.  In that discussion the 15 were eliminated from the running at the scale where $\gamma_{15}=1$ and the $\gamma_F=1$ RG scale computed from the running, giving a separation of a factor of 15. 

We can now enact this theory in the holographic model which includes the chiral symmetry breaking dynamics. 
We show the running of $\Delta m^2_{15}$  against the log of the RG scale  in Figure 1 in blue. Here we work in units of the eventual $\rho$ meson mass made of the 15 dimensional representation matter. 

Note the theory looks to have an IR fixed point but this will not be realized because chiral symmetry breaking will be triggered before it is fully reached. The 15 hits $\gamma=1$ first at a scale $\ln \mu=14.2$. This seems like a very high scale relative to the emergent $\rho$ meson mass and it is! The reason is that the IR fixed point value of the $\gamma_{15}$ is very close to $1$ (that is $\Delta m^2_{15}$ at the fixed point is just 
-1.05). Although there is an instability at a high scale we must include the full chiral symmetry breaking dynamics to see how speedily the condensate formation responds to  the instability.

In Figure 2 we show the resulting field $L_{15}(\rho)$ in the holographic model (also in blue) - it can be thought of as the 

\begin{center}
\includegraphics[width=6.7cm,height=4.8cm]{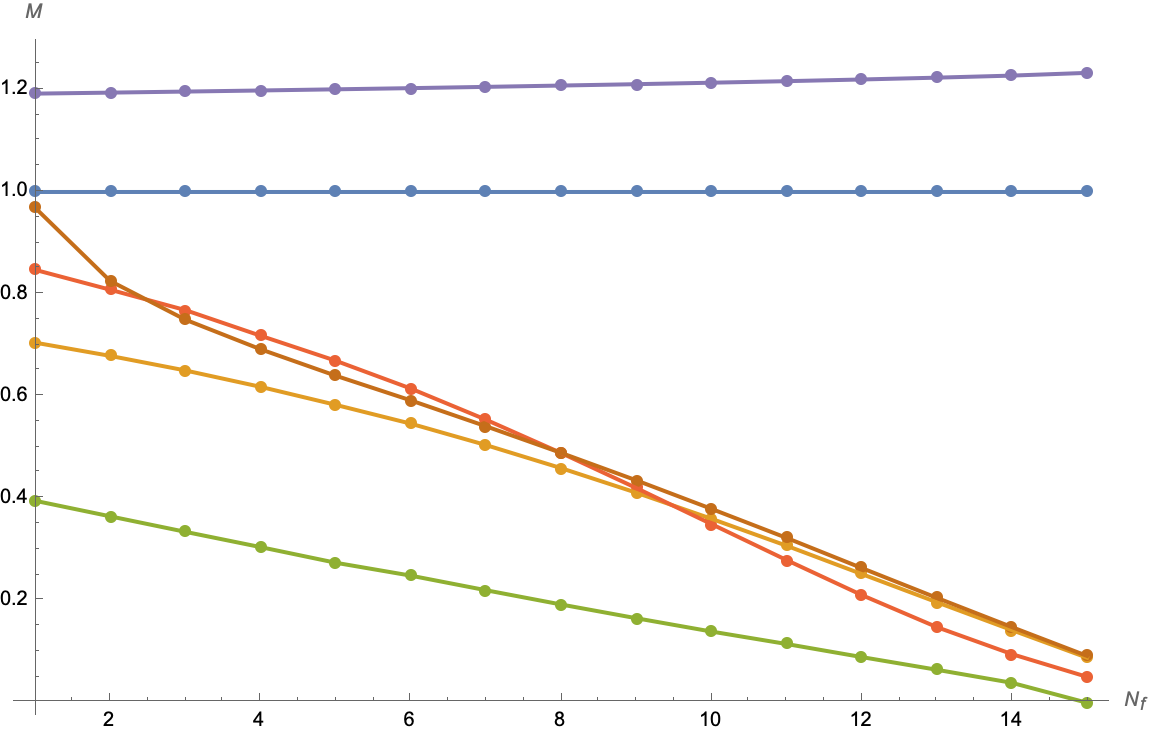}

\textit{Fig 3: Mass spectra for the SU(5) theory, $\rho$ mesons in blue (symmetric) and dark yellow (fundamental), $\sigma$ mesons in green (symmetric) and orange (fundamental), axials in purple (symmetric) and brown (fundamental). The pions in both sectors are massless at zero fermion mass.}
\end{center}

effective fermion mass as a function of RG scale. It shows chiral symmetry breaking, bending away from $L=0$ in the infra-red. As hinted at above the IR value of $L_{15}(\rho)$ is much smaller than the scale where the $\gamma=1$/BF bound is violated.

The reason for this is that the BF bound violation comes 
with a potential energy cost but, equally, derivatives of the $L(\rho)$ function also cost energy. The resultant solution is a balance between these costs. In this theory at the scale of the BF bound for the 15 the running is so slow and centred essentially on the violation value that the solution prefers to ``live with it'' rather than bending off axis immediately. Only at a much lower scale does the chiral symmetry breaking set in because the cost of the BF bound violation has built up over several orders of magnitude in RG scale. This is an example of why incorporating the chiral symmetry breaking dynamics in the model is important to understand the true mass scales. The extreme walking behaviour in the UV of this theory would make its study on the lattice quite problematic.

We now move to the fundamental sector. The running $\Delta m^2_{5}$ is shown in Figure 1 - above the scale where the 15 goes on mass-shell the running is given by the red curve. We integrate out the 15 at the IR value of $L_{15}(\rho)$ where $L_{15}(\rho)=\rho$. The running is then that of the green curve in Figure 1. It is straightforward to set up an interpolating function that incorporates the UV red running and the IR green running. $\gamma=1$ for the fundamentals is reached at $\ln \mu = -2.3$. 

We can now solve for the $L(\rho)$ function for the 5 representation. We display this in green in Figure 2. We can see there is a gap of a factor of 10.7 between the IR mass 
values for the two representations.  This is of order but not quite as large as the factor of 15 we naïvely expected. Here there is not a large gap between the BF 
bound violation scale and the emergent IR mass because

\begin{center}
\includegraphics[width=6.7cm,height=4.8cm]{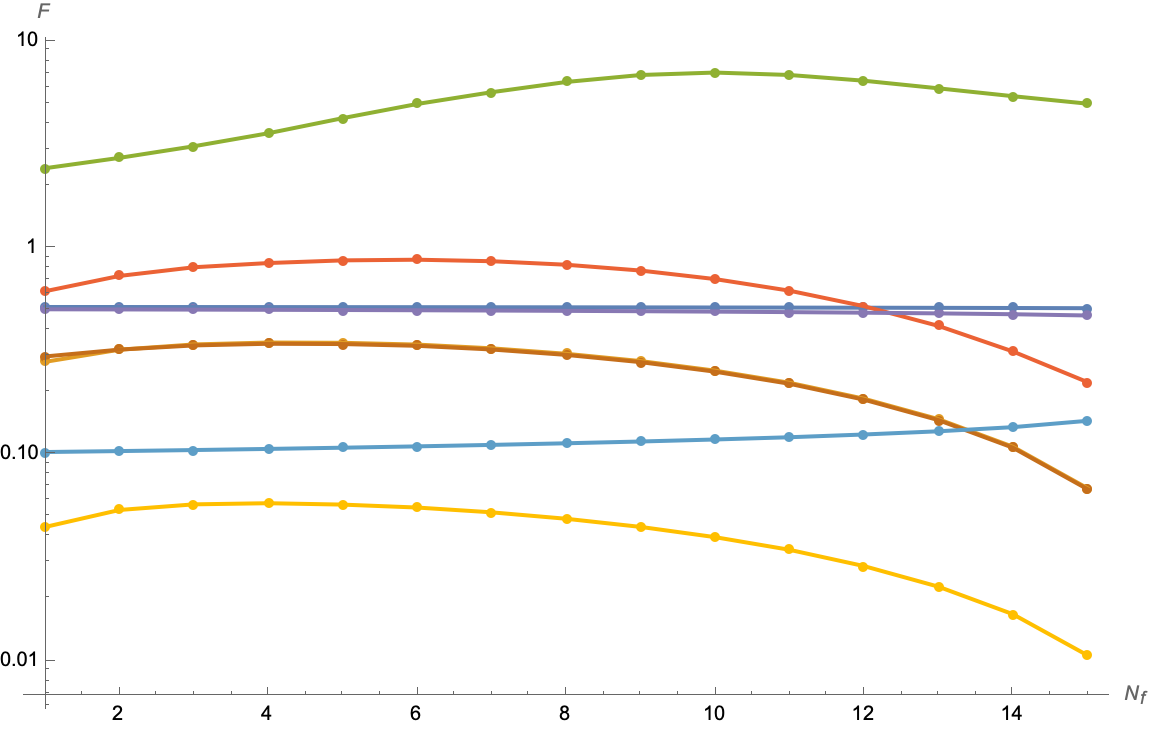}

\textit{Fig 4: Decay constants for the SU(5) theory, $\rho$ mesons in blue (symmetric) and dark yellow (fundamental), $\sigma$ mesons in green (symmetric) and orange (fundamental), axials in purple (symmetric) and brown (fundamental) and pions in cyan (symmetric) and light yellow (fundamental).}
\end{center} \vspace{-0.2cm}

the running is much faster once the 15 have decoupled (although we have chosen $N_f^F$ to slow this running to generate the gap). The gap size is therefore in line with the expectations of the running analysis of \cite{Evans:2020ztq} although both scales are significantly lower than expected from the $\gamma_{15}=1$ criteria.

It is now possible to use the holographic model to compute the spectrum of the  gauge theory. We do this not just for $N_f^F=15$ but at all $N_f^F$ values (recomputing the vacuum equivalent to Figure 2 at each value). We show the results in Figure 3. In each theory we normalize the spectrum by the $\rho$-meson mass of the 15 representation matter so the symmetric representation $\rho$ line (blue) is flat at one by construction. We display the decay constants of the theory also in Figure 4. We see that the symmetric representation $\rho$ decay constant is also essentially independent of the number of fundamental fermions (it would grow with the number of 15 representation flavours but we keep that fixed at one here).

The axial meson mass (purple) in the symmetric representation sector is larger than those of the $\rho$, though its decay constant is marginally smaller, but they are essentially unresponsive to the number of light fundamental fermions also.

The $\rho$-meson mass in the fundamental sector (dark yellow)  is a good measure of the gap. We see from Figure 3 that this $\rho$ mass falls relative to the $\rho$ mass in the 15 sector. The extreme value is a gap of $\sim$11.3 between these masses for $N_f^F=15$ (at higher $N_f^F$ the UV theory is in the conformal window).
The fundamental sector $\rho$ decay constant initially rises with $N_f^F$ but then falls as the scale 
of the gap falls at larger $N_f^F$.

The $\sigma$ meson in the symmetric representation sector (green) is the lightest state reflecting the walking in the high energy theory. Its decay constant initially grows,

\newpage

\begin{center}
\includegraphics[width=6.7cm,height=4.6cm]{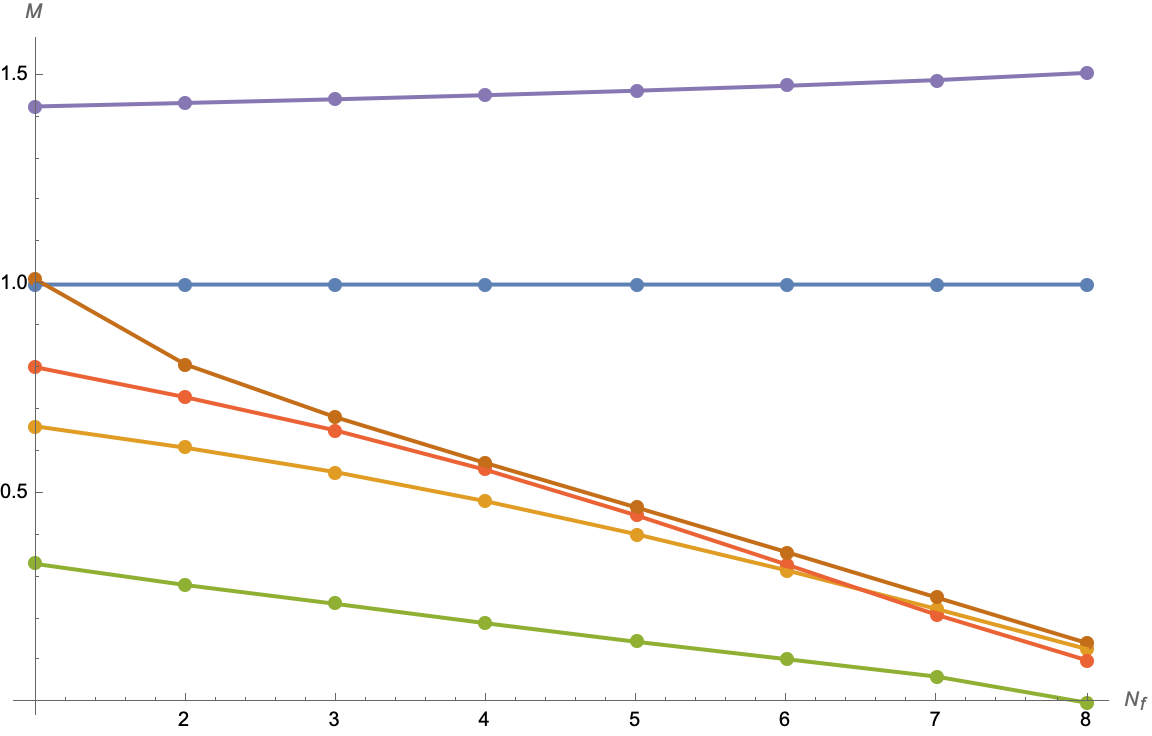}

\textit{A: Mass spectra for the SU(3) theory.}
\end{center}  \vspace{-0.75cm}

\begin{center}
\includegraphics[width=6.7cm,height=4.6cm]{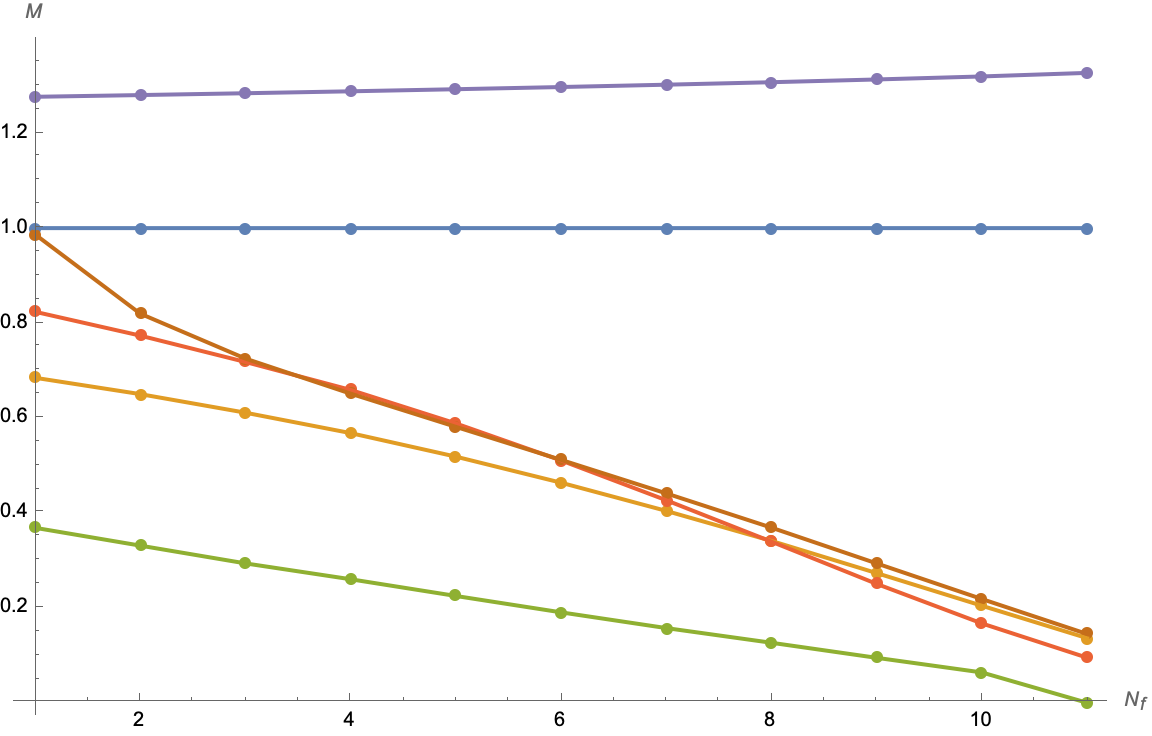}

\textit{C: Mass spectra for the SU(4) theory.}
\end{center}  \vspace{-0.75cm}

\begin{center}
\includegraphics[width=6.7cm,height=4.6cm]{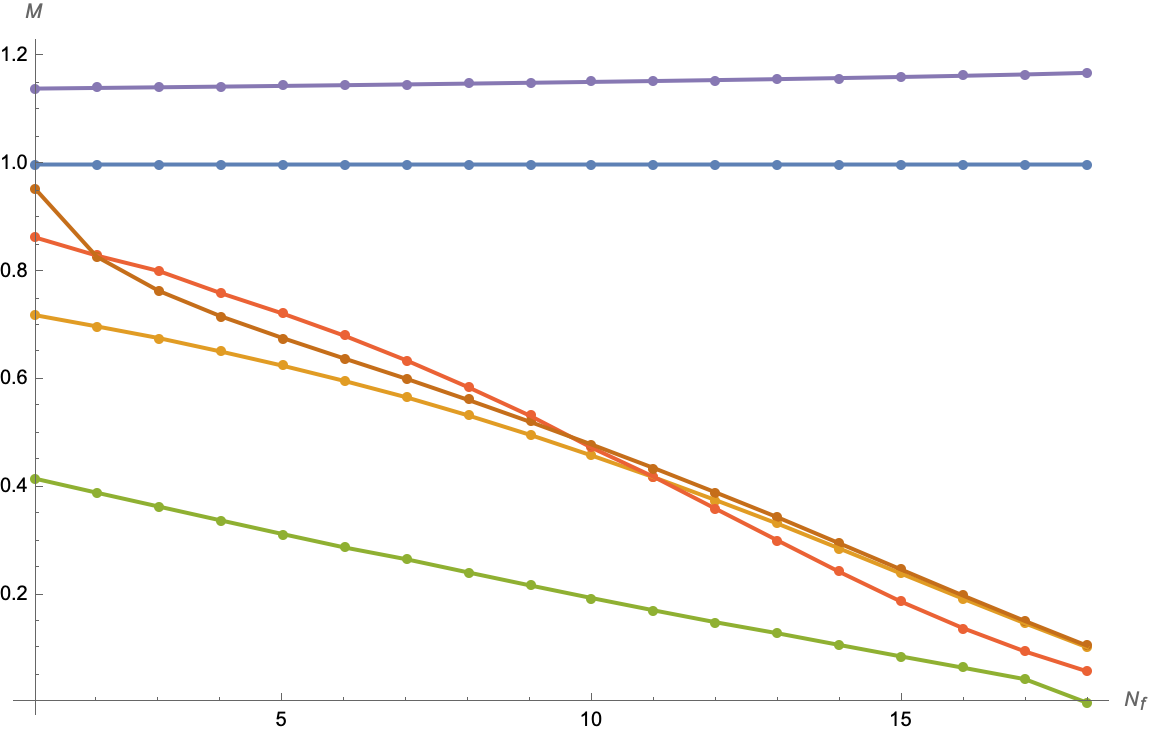}

\textit{E: Mass spectra for the SU(6) theory.}
\end{center}  \vspace{-0.75cm}

\begin{center}
\includegraphics[width=6.7cm,height=4.6cm]{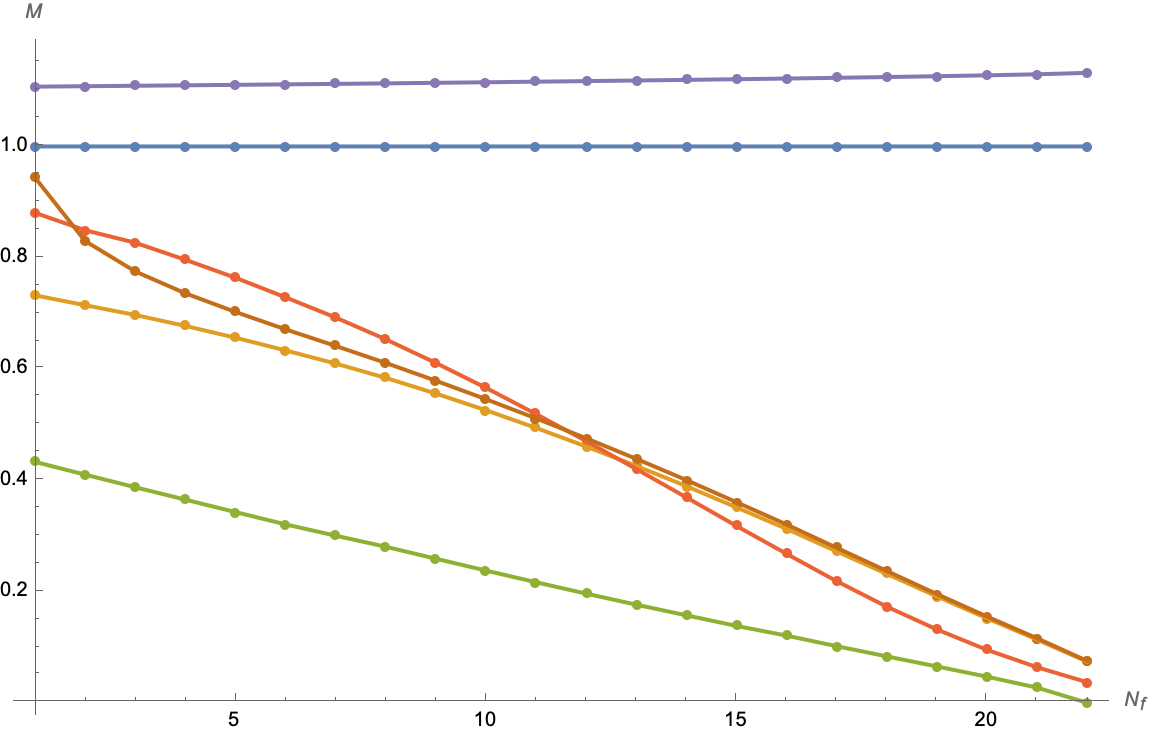}

\textit{G: Mass spectra for the SU(7) theory.}
\end{center}  \vspace{-0.25cm}

\mbox{\it Figure 5: The spectra and decay constants of the SU($N_c$) gauge theory with one two index symmetric rep. } 
\mbox{\it matter field and $N_f^F$  fundamentals for $N_c=3,4,6,7$. $\rho$ mesons in blue (symmetric) and dark yellow }
\mbox{\it (fundamental), $\sigma$ mesons in green (symmetric) and orange (fundamental), axials in purple (symmetric)}
\mbox{\it  and brown (fundamental) and pions in cyan (symmetric) and light yellow (fundamental).}

\begin{center}
\includegraphics[width=6.7cm,height=4.6cm]{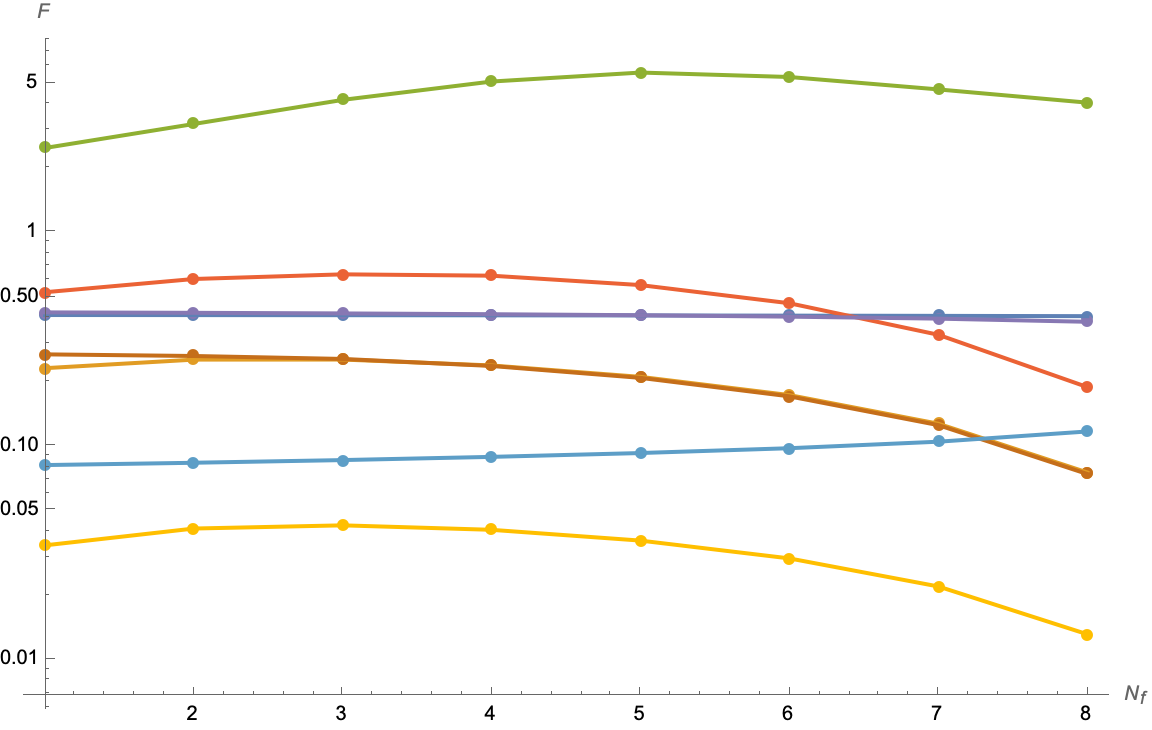}

\textit{B: Decay constants for the SU(3) theory.}
\end{center}  \vspace{-0.75cm}

\begin{center}
\includegraphics[width=6.7cm,height=4.6cm]{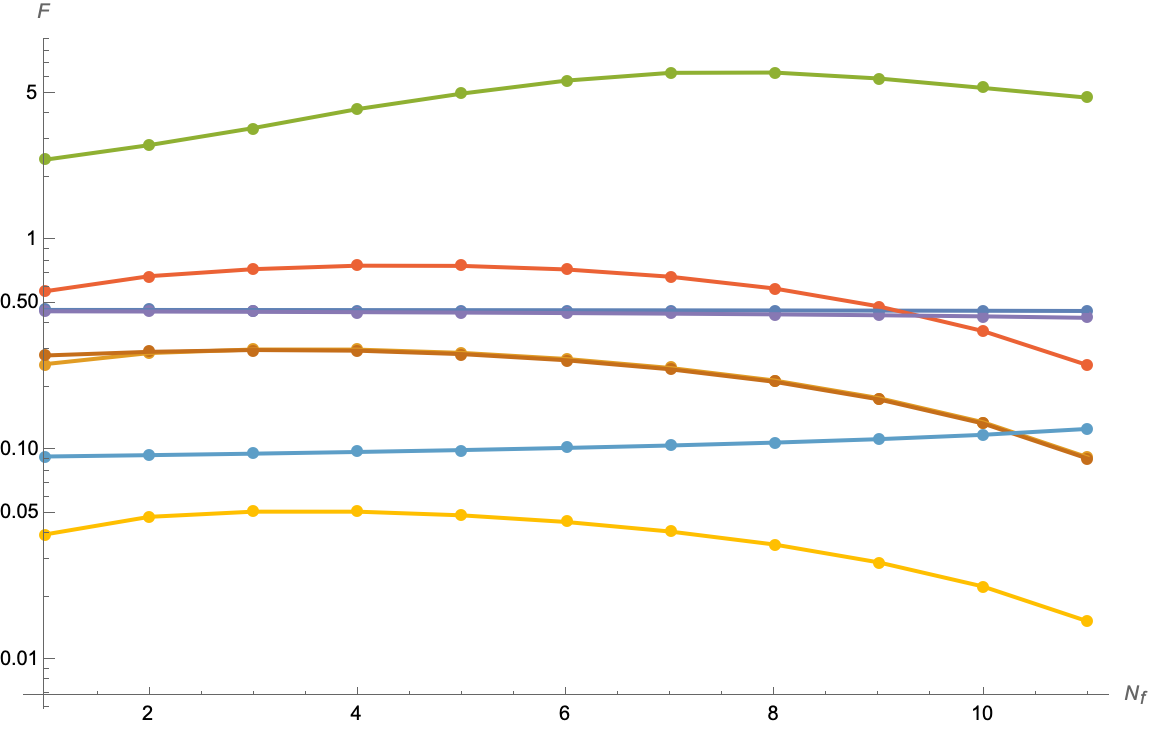}

\textit{D: Decay constants for the SU(4) theory.}
\end{center}  \vspace{-0.75cm}

\begin{center}
\includegraphics[width=6.7cm,height=4.6cm]{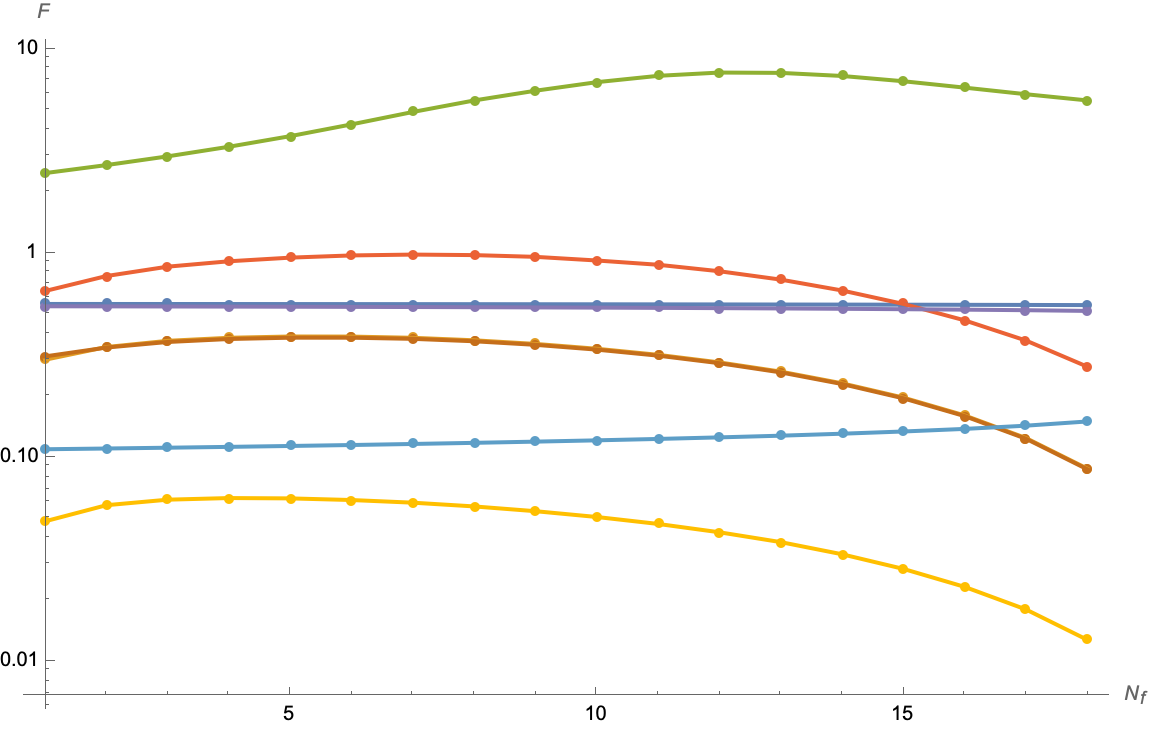}

\textit{F: Decay constants for the SU(6) theory.}
\end{center}  \vspace{-0.75cm}

\begin{center}
\includegraphics[width=6.7cm,height=4.6cm]{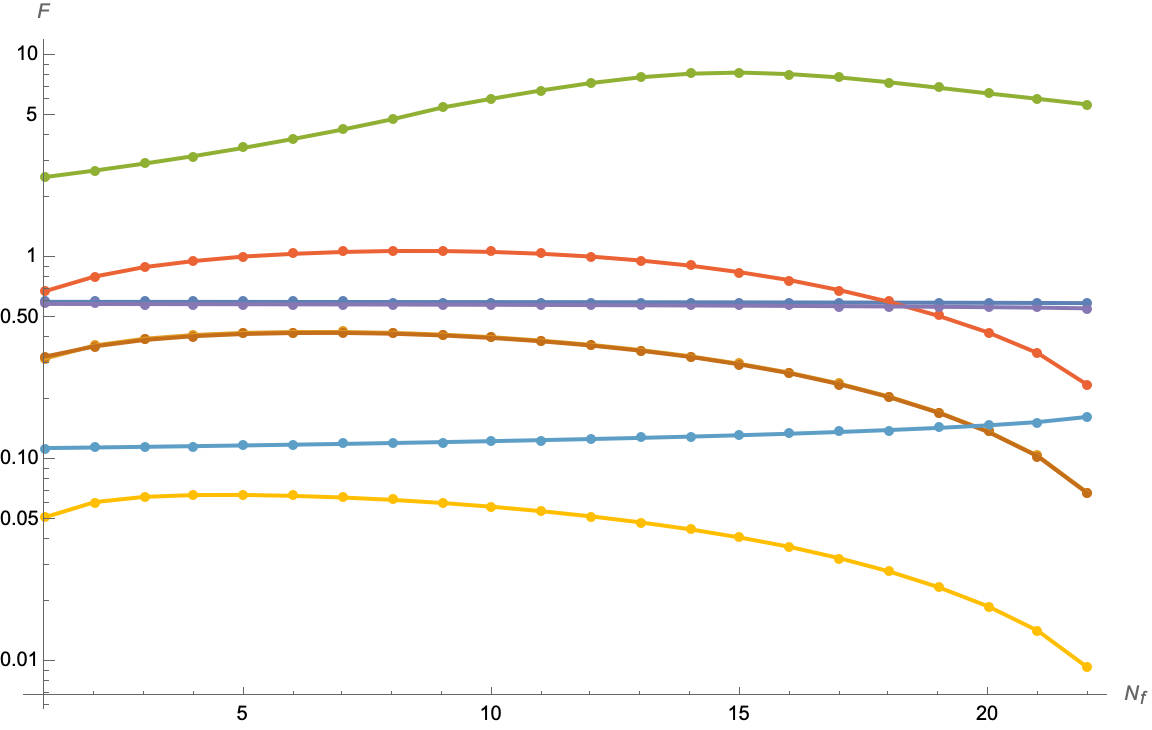}

\textit{H: Decay constants for the SU(7) theory
.}
\end{center}  \vspace{0.25cm}

\mbox{\it $\left. \right.$}

\mbox{\it $\left. \right.$}

\newpage

\begin{center}
\includegraphics[width=6.7cm,height=4.8cm]{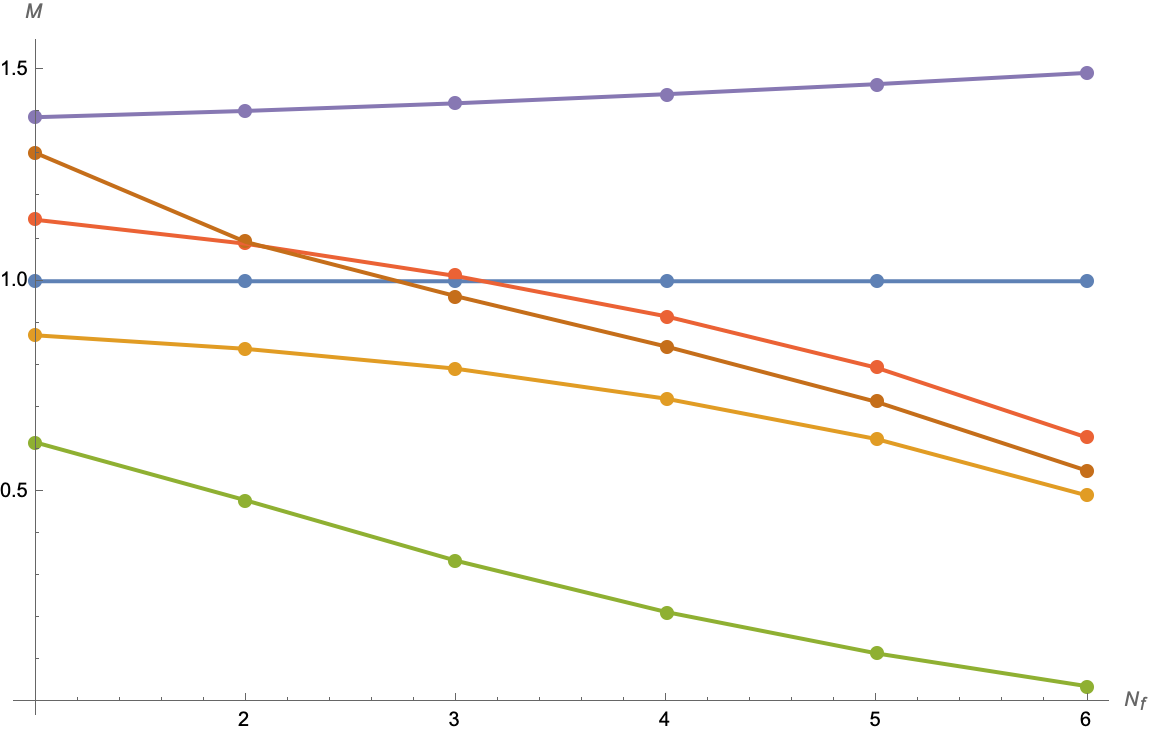}

\textit{Fig 6: Meson masses for the SU(3) theory with a factor of 2 removed in (\ref{dm}): $\rho$ mesons in blue (symmetric) and dark yellow (fundamental), $\sigma$ mesons in green (symmetric) and orange (fundamental), axials in purple (symmetric) and brown (fundamental) and pions in cyan (symmetric) and light yellow (fundamental).}
\end{center} \vspace{-0.2cm}

 before falling off at high $N_f^F$ in the more walking theory.

The $\sigma$ meson mass in the fundamental sector falls as $N_f^F$ rises as the theory below the 15's mass gap becomes also more walking. The theory is never as walking as the far UV theory though. 

Although the gap between the $\rho$ mesons in the two sectors is slightly smaller than the naïve factor of 15 that one obtains from taking the ratio of the scales at which the BF bounds are broken, it is still significant. Below the mass scale of the 15 the perturbative computation of how long the coupling takes to run to the BF bound remains the core mechanism. In principle one could study the theory at finite temperature and expect to find an intermediate phase where the 15 representation experiences chiral symmetry breaking but the fundamentals do not and presumably this phase also has no confinement.  

\subsection{Other $N_c$}

Given the gap in the SU(5) theory has not been quite as large as we expected it is interesting to look at the $N_c$ dependence of the spectrum. We have studied the $N_c=3,4,6,7$ cases as well. In each case again there is one two index symmetric representation fermion and $N_f^F$
fundamental fermions. We compute for all $N_f^F$ up to the limit in each case where our ansatz for the running of $\gamma$ places the theory in the conformal window, The results are shown in Figure 5 and are similar to the SU(5) case. The largest gaps between the $\rho$ meson masses of the two sectors are: SU(3), $N_f^F=8$: 7.34; 
    SU(4), $N_f^F=11$: 11.34;
    SU(6), $N_f^F=18$: 9.70;
    SU(7), $N_f^F=22$: 13.54.

At this point it is important to quantify at least in some way how reliable these results are. In particular they depend on the running ansatz at a given $N_f^F$ and $N_c$ and also the translation of the running $\gamma$ to $\Delta m^2$. In particular the key point is at what scale the BF bound is violated and the derivative in the running at that point. One way to test this is as discussed below (\ref{grun}) - we by hand remove the factor of 2 in (\ref{dm}) effectively doubling the critical coupling value required. This lowers the value of $N_f^F$ where the edge of the conformal window lies (to $N_f^F=6$ for example in the theory with $N_c=3$). It also narrows the gap between the two representation sectors since the coupling is running quicker in the interval below the constituent mass of the symmetric representation matter. 

As an example we show the mass spectrum for the SU(3) gauge theory in this approximation in Figure 6. The gap at $N_f^F=6$ is a factor of 2 which is not hugely different from that we saw at the same $N_f^F$ value above. However since it is the closest we can tune to the conformal window edge this result is considerably smaller than the value for the gap we had for $N_f^F=8$ previously of 7.34. As a pessimistic example it still displays a gap of a factor of 2 which is discernible and could be split by a theory with temperature between these two scales.

Returning now to the original version of (\ref{dm}), we consider  if one were to seek this behaviour on the lattice. Then it is usually simplest to study fermions in multiples of four 
$N_f^F$. One needs to be careful though, because our ansatz for where the edge of the conformal window lies need not be accurate, so one does not want to lie too close to that boundary.
The majority of the mass gap is gained only at the last few $N_f^F$ increments. Given lattice computations are easier with $N_f^F=4n$, the gaps in the following theories may be of interest:
\vspace{-0.4cm}

\begin{center}
 SU(3), $N_f^F=8$: $\frac{m_{\rho 15}}{m_{\rho F}}= 7.34$
\\
 SU(4), $N_f^F=8$: $\frac{m_{\rho 15}}{m_{\rho F}}= 2.93$
\\
 SU(5), $N_f^F=12$: $\frac{m_{\rho 15}}{m_{\rho F}} = 3.97$
\\
 SU(6), $N_f^F=16$: $\frac{m_{\rho 15}}{m_{\rho F}}= 5.18$
\\
 SU(7), $N_f^F=20$: $\frac{m_{\rho 15}}{m_{\rho F}}= 6.59$
\end{center}\vspace{-0.4cm}

The example of SU(5) with $N_f^F=15$ though, warns that the UV of these theories might be very walking with widely separated scales. That would  make analysis hard on the lattice. 

In figure 7 we show the running of $\Delta m^2$ for each case. In all cases the running in the UV theory is faster at the symmetry breaking scale than the case in Figure 1. The  scales of BF bound violation in the two sectors are more compressed.In fact the SU(4) theory has the strongest 
running at the 15 condensation scale and so may be the easiest to study on the lattice to find a gap in scales. Of course in that case the gap is only a factor of three - but this trade off may ease the computation.

\begin{center}
\includegraphics[width=6.7cm,height=4.8cm]{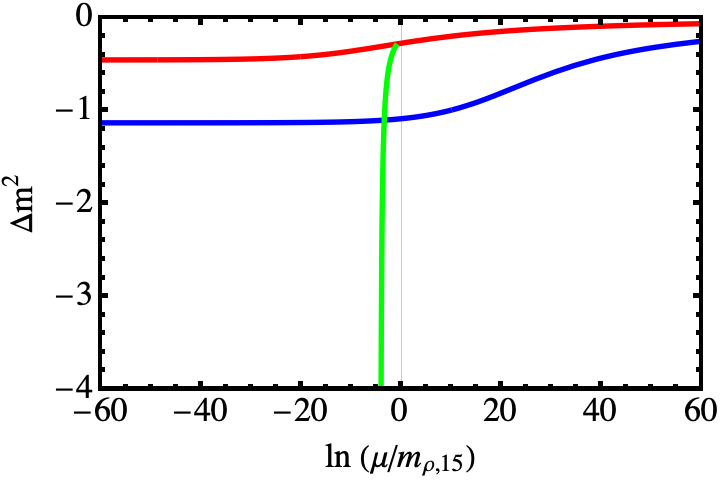}
\vspace{-0.2cm}

\textit{A: SU(3) with $N_f^F=8$: $\Delta m^2=-1$ occurs at 9.3 for the 6 and
-2.1 for the 3 with the 6 decoupled.}
\end{center} \vspace{-0.2cm}

\begin{center}
\includegraphics[width=6.7cm,height=4.8cm]{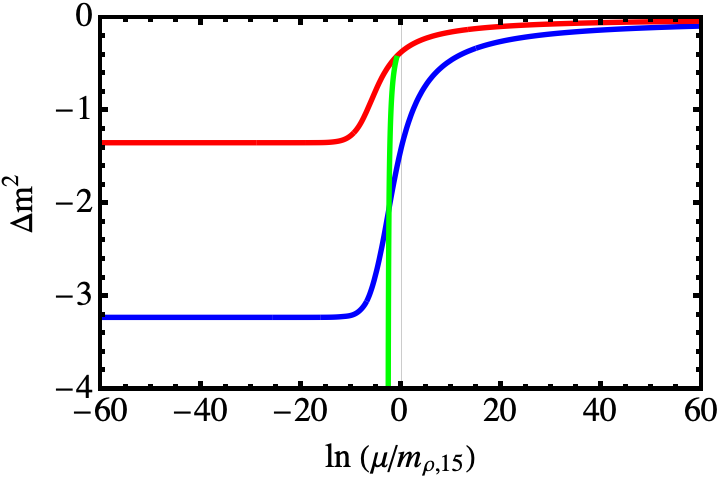}

\textit{B: SU(4) with $N_f^F=8$: $\Delta m^2=-1$  occurs at 2.2 for the 10 and
-1.3 for the 4 with the 10 decoupled.}
\end{center} \vspace{-0.2cm}

$\left. \right.$ \hspace{5cm} \includegraphics[width=6.7cm,height=4.8cm]{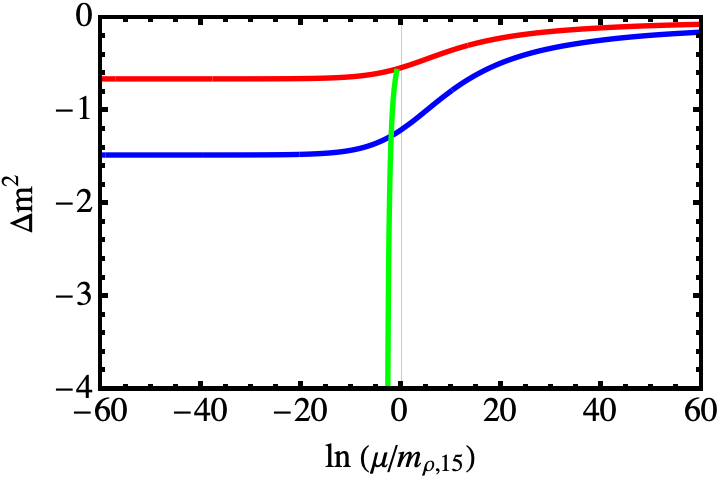}\vspace{-0.2cm}

$\left. \right.$ \hspace{5cm}  \mbox{\textit{E: SU(7) with $N_f^F=20$:$\Delta m^2=-1$ occurs at 4.9 }}
$\left. \right.$ \hspace{5cm}  \mbox{\textit{for the 28 and
-1.8 for the 7 with the 28 decoupled. }}
\vspace{-0.2cm}

 \mbox{\it Fig 7: The running of $\Delta m^2$ vs $\ln \mu$, for theories with $N_f^F$ multiples of 4 that give the largest spectral gaps, for}
\mbox{\it the two index symmetric  rep.  (blue), fundamental (red) and for the fundamentals with the higher}
\mbox{\it dim . rep. decoupled below the scale where it is  on mass shell (green). The energy scales are given in units of}
\mbox{\it the $\rho$-meson mass in the $15$ sector.}

\begin{center}
\includegraphics[width=6.7cm,height=4.8cm]{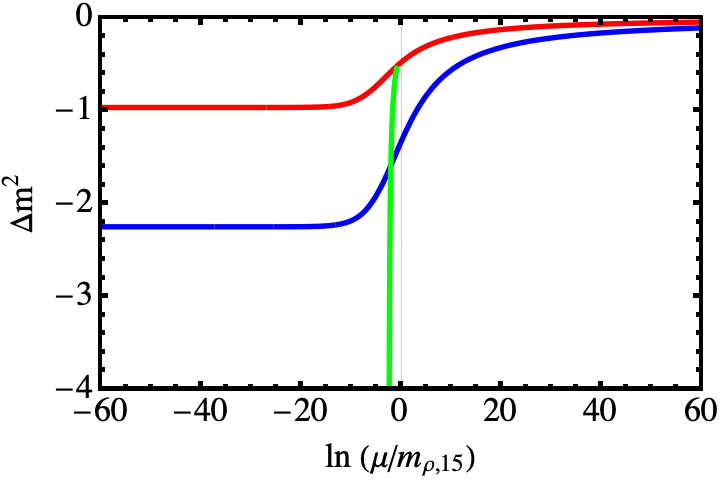} \vspace{-0.2cm}

\textit{C:  SU(5) with $N_f^F=12$: $\Delta m^2=-1$ occurs at 3.0 for the 15 and
-1.5 for the 5 with the 15 decoupled.}
\end{center} \vspace{1.5cm}

\begin{center}
\includegraphics[width=6.7cm,height=4.8cm]{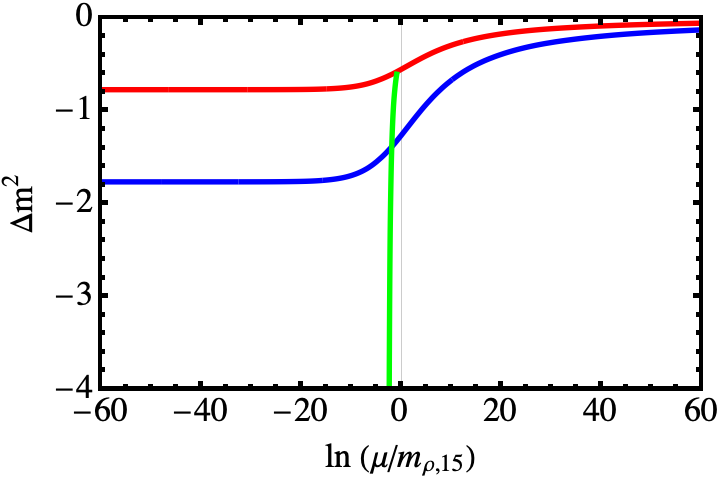} \vspace{-0.2cm}

\textit{D:  SU(6) with $N_f^F=16$: $\Delta m^2=-1$  occurs at 3.8 for the 21 and
-1.7 for the 6 with the 21 decoupled.}
\end{center}\vspace{-0.6cm}

\newpage

\begin{center}
\includegraphics[width=6.7cm,height=4.8cm]{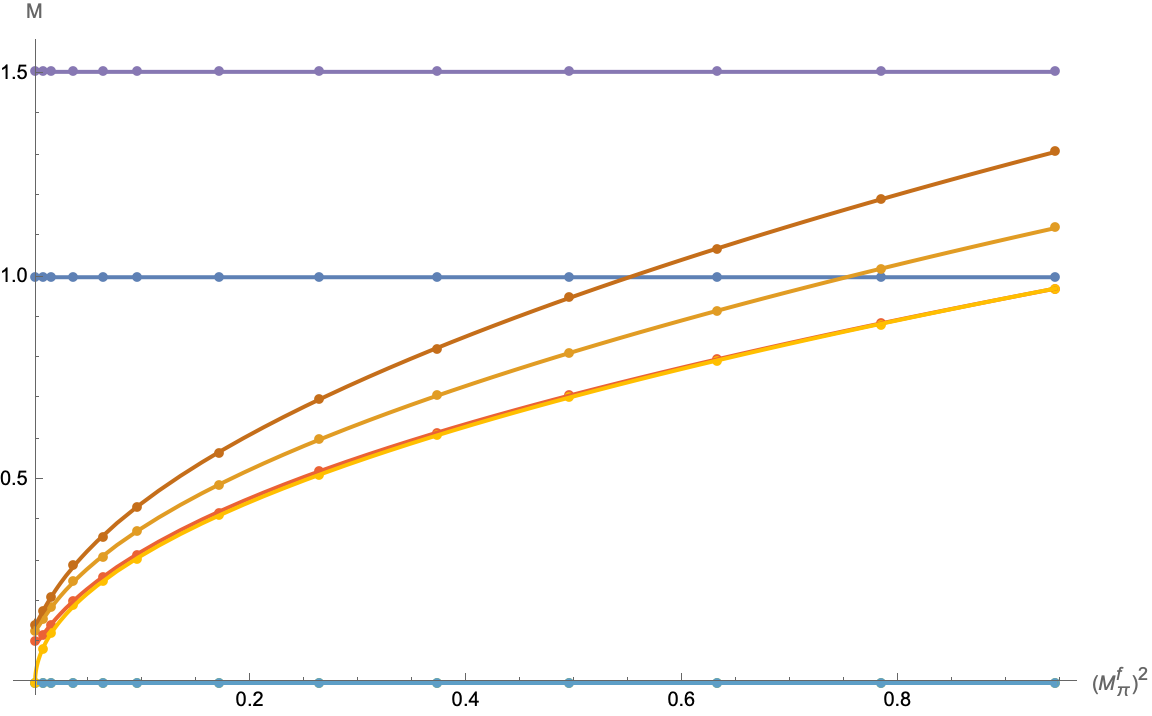}

\textit{Fig 8: Mass spectra for the SU(3) theory with $N_f=8$ and non-zero UV mass. $\rho$ mesons in blue (symmetric) and dark yellow (fundamental), $\sigma$ mesons in green (symmetric) and orange (fundamental), axials in purple (symmetric) and brown (fundamental) and pions in cyan (symmetric) and light yellow (fundamental).}
\end{center}  

\section{A Massive Theory}

Finally we explore briefly the dependence of the gap on the fundamental quark mass. This mass is the most crucial to keep small to maintain the gap size. We study this mass dependence in Fig 8 and Fig 9 for the SU(3) theory with $N^F_f=8$ showing  the spectrum's dependence on the pion mass in the fundamental sector (which is proportional to the square root of the quark mass at small quark mass). Clearly to maintain the gap one will need the quark mass typically an order magnitude smaller than the chiral symmetry breaking scale of the higher representation. 

\section{Conclusions}

We have used a holographic model to study the spectrum of $SU(N_c)$ gauge theories with one Dirac fermion in the two index symmetric representation and $N_f^F$ Dirac  fermions in the fundamental representation.  The model predicts the chiral symmetry breaking scale and meson

\begin{center}
\includegraphics[width=6.7cm,height=4.8cm]{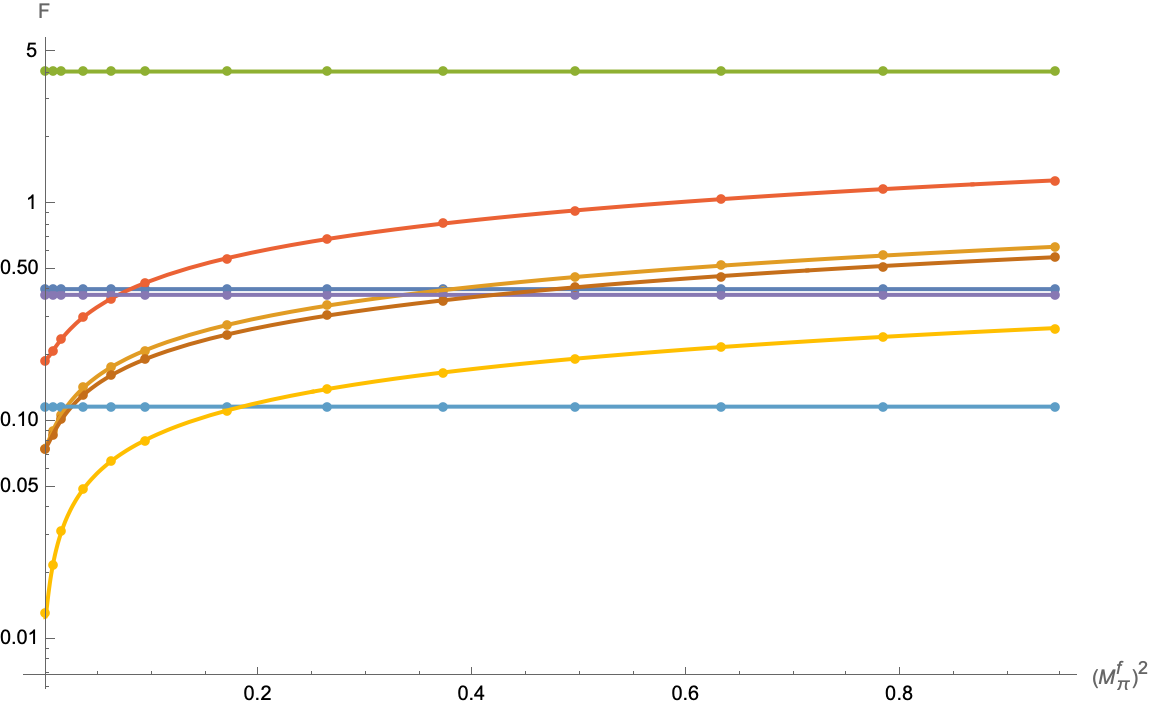}

\textit{Fig 9: Decay constants for the SU(3) theory with $N_f=8$ and non-zero UV mass.  $\rho$ mesons in blue (symmetric) and dark yellow (fundamental), $\sigma$ mesons in green (symmetric) and orange (fundamental), axials in purple (symmetric) and brown (fundamental) and pions in cyan (symmetric) and light yellow (fundamental).}
\end{center}  

spectrum once the dynamics is included through a running anomalous dimension. We have extrapolated the
 perturbative two loop runnings into the non-perturbative regime to provide estimates of these running $\gamma$. We have
identified a number of theories with spectral gaps between the two $\rho$ mesons made of fundamental and two index symmetric matter. These gaps could potentially be larger than an order of magnitude. Even in the case where we used a very pessimistic ansatz for the critical coupling value the gaps were still discernible and as large as a factor of two.  It would be interesting to study these theories on the lattice to confirm these gaps and to realize theories at finite temperature with chiral symmetry breaking but potentially no confinement.  Many of the theories with the largest gaps are quite walking with separated scales but we have identified less walking theories that have a gap that may be easier to study. For example SU(4) gauge theory with $N_f^F=8$ may be a candidate with the smaller gap of three but more QCD-like running. \vspace{-0.5cm}

\bigskip \noindent {\bf Acknowledgements:}  
NE's work was supported by the
STFC consolidated grants ST/T000775/1 and ST/X000583/1.


\end{document}